\documentclass{article}
\usepackage{PRIMEarxiv}

\usepackage[utf8]{inputenc} % allow utf-8 input
\usepackage[T1]{fontenc}    % use 8-bit T1 fonts
\usepackage{hyperref}       % hyperlinks
\usepackage{url}            % simple URL typesetting
\usepackage{booktabs}       % professional-quality tables
\usepackage{amsfonts}       % blackboard math symbols
\usepackage{nicefrac}       % compact symbols for 1/2, etc.
\usepackage{microtype}      % microtypography
\usepackage{lipsum}
\usepackage{fancyhdr}       % header
\usepackage{graphicx}       % graphics
\graphicspath{{media/}}     % organize your images and other figures under media/ folder
\usepackage[
    backend=biber,
    style=ieee,
    sorting=none
]{biblatex}
\usepackage{quantikz}
\usepackage{listings}

\lstdefinelanguage{Qasm}{
  morekeywords=[1]{gate,measure,reset,if,else,include,opaque,
    qreg,creg,OPENQASM},
  morekeywords=[2]{x,y,z,h,s,sdg,t,tdg,rx,ry,rz,cz,cx,ccx,swap,
    u1,u2,u3,cu1,cu2,cu3,barrier,ctrl,reset},
  morekeywords=[3]{true,false,pi},
  sensitive=true,
  morecomment=[l]{//},
  morecomment=[s]{/*}{*/},
  morestring=[b]",
  alsoletter={_},
}
\title{A machine learning architecture for scaling of holographic quantum circuits}

\author{Cliff Sun \& Michael L. Wall \\
Johns Hopkins Applied Physics Laboratory \\
Laurel, MD 20723, USA 
}

\begin{document}
\maketitle

\begin{abstract}
We describe an architecture and workflow for learning and generalizing the syntax and parameters of \emph{holographic quantum circuits} (HQCs), which are natural quantum circuit analogs of classical tensor networks.  An HQC uses mid-circuit measurements and resets on a small subset of ``physical'' qubits whose measurement outcomes become correlated through coupling with a register of persistent ``ancilla" qubits, analogous to the physical and bond indices of a tensor network.  Hence, an HQC with fixed qubit resources can efficiently generate samples from a large quantum state, and adding qubits to the HQC ansatz has the effect of exponentially increasing the effective bond dimension.  Our machine learning approach provides mechanisms for harnessing both features of HQCs by extrapolating the syntax (e.g., a sequence of gates, measurements, and resets) and parameters (e.g., the angles appearing in single-qubit rotations) from a dataset of HQCs to longer or more expressive HQCs. We demonstrate the methodology and its performance on an analytically solvable dataset formed by a vector-valued generalization of the W-state. Finally, we provide an outlook for usage of the methodology for scaling the effective bond dimensions of HQCs in quantum simulation.
\end{abstract}
\section{Introduction}

Over the past decade, quantum computing has seen a rapid surge of progress, driven by the prospect of exponential speed-ups in a variety of domains such as optimization \cite{grover-algorithm}, cryptography \cite{BENNETT20147}, and high-dimensional many-body simulations \cite{fauseweh-2024}. At the same time, advances in scaling of qubit resources and fidelity \cite{lall2025reviewcollectionmetricsbenchmarks} coupled with the development of feasible quantum error correction protocols \cite{collaborators-2024} have dramatically reshaped the landscape of which is experimentally feasible. Nevertheless, present-day devices remain in the noisy-intermediate-scale-quantum (NISQ) regime \cite{preskill-2018}, where limited qubit connectivity, short coherent times, and imperfect gate operations impose severe constraints on algorithmic performance. A central focus in the NISQ regime has been on variational quantum circuits, in which a classical representation of the quantum circuit is optimized based on feedback from a quantum device. Within this paradigm, one means of moving towards the regime of quantum advantage is to scale the circuit to where it cannot be simulated directly using classical means. A central obstacle to reaching this scale is the lack of an efficient quantum-circuit design strategy that can autonomously learn and scale circuits for simulations with classically-intractable sizes.  In this work, we develop such a pipeline utilizing techniques from tensor networks and machine learning, demonstrate its effectiveness in an analytically tractable scenario, and present an outlook for applying it to the problem of scaling variational quantum circuits.

% Variational quantum circuits
Classical tensor networks (CTNs) describe a large-rank tensor in terms of a contracted network of small-rank tensors, and can realize exponential improvements in storage and computation time compared to a dense tensor representation when the maximum dimension of any contracted index - known as the \emph{bond dimension} - remains small.  CTNs, in particular the one-dimensional CTN structure known as a Matrix Product State (MPS), have had a far-ranging impact on computational quantum science.  They have enabled deeper insight into the entanglement and correlation structure of many-body quantum states, as well as forming the underpinning of the most accurate numerical methods for strongly correlated low-dimensional systems, such as the density matrix renormalization group (DMRG)~\cite{SCHOLLWOCK201196}.  

Beyond classical tensor representations, tensor networks also provide a robust framework for designing parameterized quantum circuits that are resource efficient and implementable with existing quantum hardware.  A natural generalization of CTNs to quantum circuits is provided by holographic quantum circuits (HQCs)~\cite{PhysRevResearch.3.033002,PhysRevA.105.062439}. HQCs are quantum circuits who utilize mid-circuit measurements and resets on a subset of ``physical" qubits, and whose measurement outcomes are correlated through a fixed register of ``ancilla" (virtual) qubits, with direct analogies to the physical and bond indices of a CTN. In its most prevalent instantiation, there is a single physical qubit that is measured and reset after each multi-qubit operation with the ancilla register, which reproduces the well-known sequential preparation scheme of an MPS \cite{PhysRevA.75.032311}. In this scheme, a single physical qubit suffices to produce samples of an arbitrarily large quantum state, and adding qubits to the HQC ansatz exponentially increases its effective bond dimension.

\begin{figure*}[t]
\centering
\includegraphics[width=\linewidth]{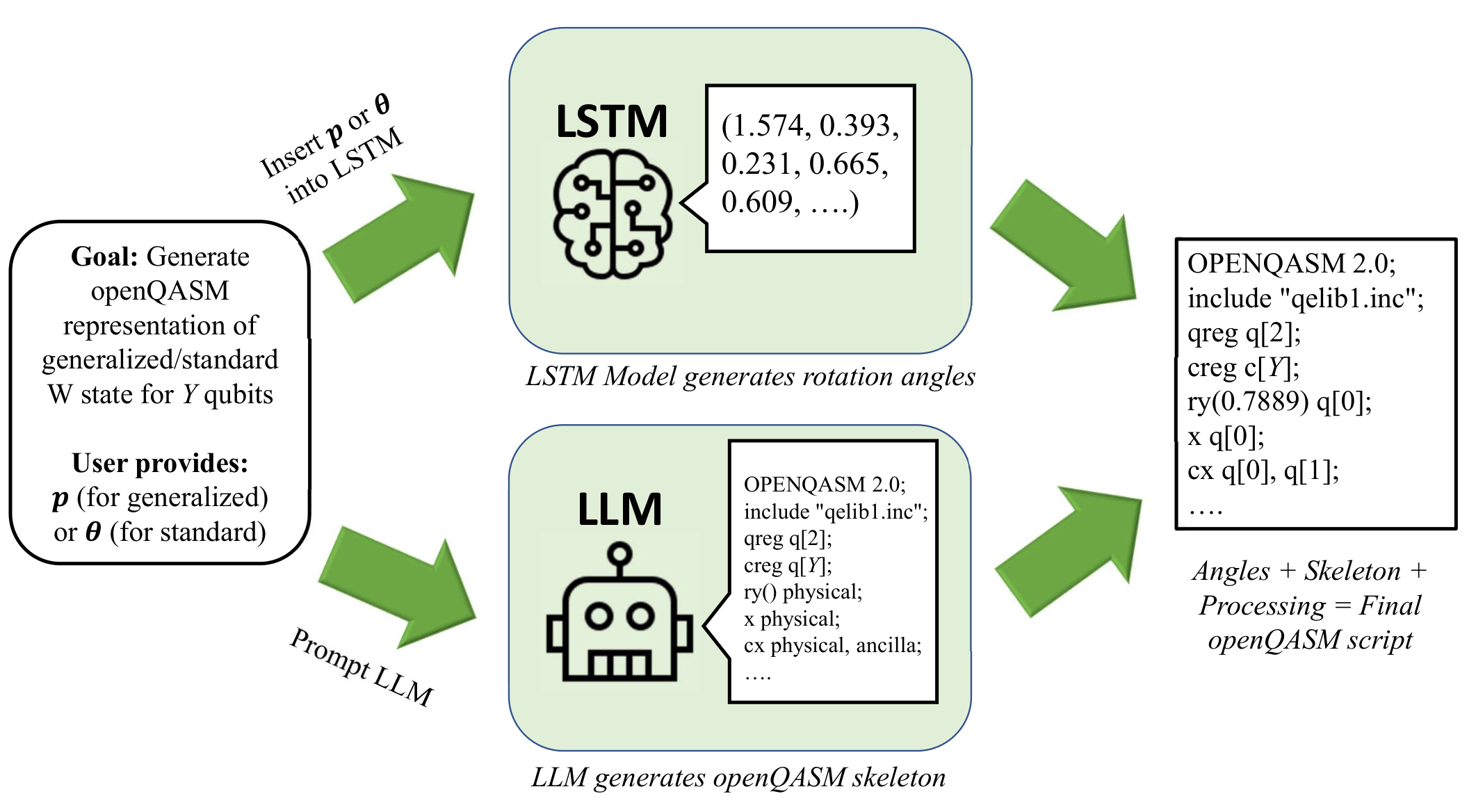}
\caption{\textit{Overview of holographic quantum circuit (HQC) generation workflow.} The user supplies a parameter vector (e.g., a classical representation of the state $\mathbf{p}$ or a collection of quantum circuit parameters $\boldsymbol{\theta}$), which is fed into the LSTM model that predicts the required rotation angles for an HQC of $Y$ qubits. In parallel, a Large Language Model (LLM) generates a "skeleton" of the quantum circuit syntax (e.g., in openQASM). The angle data from the LSTM and the openQASM skeleton from the LLM are merged and post-processed, yielding a complete quantum circuit that directly implements the desired W state on a quantum processor.}
\label{fig:workflow}
\end{figure*}

While HQCs have a clear theoretical scaling advantage over CTNs, methods to directly optimize an HQC on quantum hardware remain much less developed than very robust classical optimization techniques for CTNs like DMRG.  The naive approach of starting from a random HQC with random parameters is known to suffer from the ``barren plateau" phenomenon, in which the parameter landscape becomes featureless as its dimensionality scales.  However, the development of techniques to "compile" an optimized CTN into a HQC for given target hardware provides an opportunity to develop datasets of exemplar HQCs that can then potentially be reasoned over or extrapolated using machine learning (ML), generating new trial HQCs that may be much closer to optimality, and thereby mitigating the barren plateau problem.

ML, in addition to transforming many areas of modern science and industrial practice, has brought forth promising applications for efficient quantum circuit generation. Previous studies have shown that ML architectures such as Reinforcement Learning \cite{zhu-2023, paris-2021} and Transformers \cite{daimon-2024} can efficiently design quantum circuits while preserving high state fidelity. However, these approaches don't directly address the needs of an ML architecture for scaling up HQCs to larger numbers of qubits.  Namely, as a quantum state's dimension is increased, building out its HQC representation requires one to keep track of the following: (1) where to appropriately place each quantum gate, (2) which qubit(s) (physical and/or ancilla) each gate acts on, and (3) when and where to perform measurement and reset operations on the physical qubit all while (4) jointly optimizing the parameters of all single-qubit gates.

In Fig.~\ref{fig:workflow}, we display our proposed hybrid ML architecture used to efficiently scale up HQCs. Our approach employs a Long Short-Term Memory (LSTM) \cite{hochreiter-1997} model and a fine-tuned pre-trained Large Language Model (LLM) with Chain of Thought (CoT) \cite{wei2023chainofthoughtpromptingelicitsreasoning} to individually learn and extrapolate each respective feature of an HQC: (1) parameters (e.g. single qubit rotation angles) and (2) syntax (e.g. sequence of gates, measurements, and resets). Then, by inserting the predicted rotation angles into the scaled-up syntax, we obtain a final scaled-up HQC.  We stress that in our hybrid ML architecture, the LSTM works strictly with numerical data and the LLM works strictly with syntax, of which we represent in open Quantum Assembly Language (openQASM)~\cite{cross2017open}. As we will demonstrate, our architecture is capable of learning from a dataset of HQCs up to $n$ qubits, then scaling up to HQCs with $m>n$ qubits.

It may be argued that perhaps a fine-tuned, pre-trained LLM could be applied towards scaling-up HQCs by simultaneously extrapolating its rotation angle parameters and syntax. While LLMs have seen enormous success in learning natural language syntax and grammatical rules \cite{vaswani-2017}, they notoriously struggle with numerical data \cite{zhao2026largelanguagemodelsbad, li2025exposingnumeracygapsbenchmark, mirzadeh2025gsmsymbolicunderstandinglimitationsmathematical}. Numerical data is critical for defining the single qubit rotation angles in a quantum circuit and avoiding barren plateaus, and any misrepresentation of the rotation angle parameters can significantly change the final quantum state. Therefore, it is natural to decompose the challenge of scaling HQCs into the individual scaling of parameters and syntax, in which separate specialized ML models are assigned to generalize the facets individually. 

We benchmark the methodology and performance of our hybrid ML architecture on an analytically solvable dataset formed by a vector-valued generalization of the W state, a well-known entangled state from quantum information theory.  The W-state (and our vector-valued generalization) also has the additional property that its minimal bond dimension MPS representation is \emph{not} translationally invariant even when the state itself is, in contrast to many other states with known MPS representations like the GHZ state~\cite{perezgarcia2007matrixproductstaterepresentations}.  Because of this property, the generalization of W-state HQC representations to longer circuits is non-trivial.  We demonstrate that our architecture is capable of scaling up the generalized W state by learning and generalizing an n-dimensional vector-to-angles mapping ($\mathbf{p}_n \rightarrow \boldsymbol{\theta}_n$) and the standard W state by extending rotation angles corresponding to an $n$-qubit state to rotation angles of $m>n$ qubit state ($\boldsymbol{\theta}_n \rightarrow \boldsymbol{\theta}_{m}$).

The paper is organized as follows. In Section ~\ref{sec:w_decomp}, we present an analytical construction of the generalized W state with holographic algorithms. In Section~\ref{sec:ML_arch}, we present and justify our proposed hybrid ML architecture used to scale up HQCs. In Section~\ref{sec:benchmark}, we benchmark our hybrid ML architecture on scaling up the HQC representation of the generalized W state and the standard W state. Finally, we conclude the paper by providing an outlook for the application of our proposed hybrid ML architecture towards achieving large scale quantum simulations.

\section{\label{sec:w_decomp}Analytical Holographic Compilation of the generalized W state}

In this section, we present our vector-valued generalization of the W state and an analytic construction of holographic algorithms for its construction on two qubits.  Namely, the generalized W state on $n$ qubits is defined as
\begin{align}
|W\left(\mathbf{p}\right)\rangle&=\sum_{i=1}^{n}\sqrt{p_i}|00\dots 1_i\dots 00\rangle\, , 
\end{align}
in which $\mathbf{p}$ is a vector of positive real numbers whose elements sum to 1. Measuring $|W(\mathbf{p})\rangle$ in the computational basis yields the
one‑hot outcome $|0\ldots01_i0\ldots0\rangle$ with probability $p_i$.
When $\mathbf{p}=\mathbf{1}/n$ the state reduces to the standard $W$ state. Such a state permits an MPS representation with bond dimension $2$. After gauging in left canonical form, the isometries of the MPS are then:  

\begin{align}
   \label{eq:Liso} \mathbb{L}^{[j]i_j}&=\left(\begin{array}{cc} \delta_{i_j,0}&\sin\theta_j\delta_{i_j,1} \\ 0&\cos\theta_j\delta_{i_j,0}\end{array}\right)\, ,\\
    \sin\theta_j&=\sqrt{\frac{p_j}{\sum_{i\le j}p_i}}\, ,\cos\theta_j=\sqrt{\frac{\sum_{i<j}p_i}{\sum_{i\le j}p_i}}\, .
\end{align}

Because each $\mathbb{L}^{[j]i_j}$ is an isometry, it can be embedded into a unitary $\hat{U}_L^{[j]}$ acting on a physical qubit and an ancilla qubit:

\begin{align}
L^{[j]i}_{\alpha \beta} = \langle i\alpha |\hat{U}_L^{[j]}|0\beta\rangle\, ,
\end{align}

Using these unitaries, the generalized W state can be constructed as
\begin{align}
    \label{eq:Wdecouple}\left(\hat{U}^{[1]}_L |0_1\rangle \dots \left(\hat{U}_L^{[n-1]} |0_{n-1}\rangle \left(\hat{U}_L^{[n]} |0_n 0\rangle\right)\right) \dots\right)&=|W\left(\mathbf{p}\right)\rangle|0\rangle\, .
\end{align}

Consequently, this construction can be interpreted as a sequential quantum circuit acting on a single physical qubit and a single ancilla qubit. The explicit derivation and gate decomposition of $\hat{U}_L^{[j]}$ for $j <n$ and for $j = n$ is given in Appendix~\ref{apdx:derivation}.

\section{\label{sec:ML_arch}Hybrid Machine Learning Framework}

In this section, we describe in detail how our hybrid machine learning framework applied towards scaling holographic quantum circuits (HQCs). Here, the task of scaling HQCs is divided into two parts: (1) scaling parameters and (2) scaling quantum circuit syntax (represented in openQASM). To tackle (1), we implemented a Long Short-Term Memory (LSTM) model \cite{hochreiter-1997}, and to tackle (2), we implemented a fine-tuned pre-trained Large Language Model (LLM) with Chain of Thought (CoT) \cite{wei2023chainofthoughtpromptingelicitsreasoning}. Below, we explain and justify the application of these models towards these scaling problems.

\subsection{Parameter Generalization}

The problem of parameter generalization can be decomposed into two cases. 

\textbf{Case 1:} Suppose an \(n\)-qubit quantum state is described by a classical vector 
\(\mathbf{p}\) that is mapped to a vector of
circuit parameters (e.g., single-qubit rotation angles) \(\boldsymbol{\theta}\).  A learning model
trained on the mapping \(\mathbf{p}\rightarrow \boldsymbol{\theta}\) for
\(n\) qubits should be able to extrapolate this relationship to quantum
states with \(m>n\) qubits, i.e., to produce the appropriate
rotation‑angle vector for any \(m>n\). In other words, the model learns the mapping $f_p$: 

\begin{equation}\label{eq:p_to_theta}
    f_p: \mathbf{p} \rightarrow \boldsymbol{\theta}
\end{equation}

Such a problem is applicable to the generalized W state, whose state is parameterized by a positive, real-valued probability vector $\mathbf{p} \in \mathbb{R}^n_+$ with $\sum_i p_i = 1$. 

\textbf{Case 2:} Suppose an $n$-qubit quantum state is parameterised solely in terms of its quantum circuit parameters \(\boldsymbol{\theta}\) within a defined circuit architecture. Then, the model’s task is to take an input \(\boldsymbol{\theta}\) that corresponds to an
\(n\)-qubit state and correctly extend it to a parameter vector that
describes a state with a larger number of qubits (\(m>n\)). In other words, the model learns the mapping $f_\theta$:

\begin{equation}
    f_\theta: \mathbb{R}^{n} \ni \boldsymbol{\theta} \rightarrow \boldsymbol{\theta} \in \mathbb{R}^{m}
\end{equation}
% with $\mathbf{p}\propto \mathbf{q}$

This problem is applicable to the standard W state because there is an unambiguous definition of the state for any $n$.  For the general W state with arbitrary $\mathbf{p}$, the scaling of $\mathbf{p}$ to a longer sequence is ambiguous, and so a direct learning of the scaling up process is ill-defined.

We treat both cases of parameter generalization as a time series forecasting problem, since the known part of the sequence (either $\boldsymbol{\theta}$ or $\mathbf{p}$) can be used to predict its continuation ($\boldsymbol{\theta}$ for larger qubit counts). Recurrent neural networks (RNNs) are a natural choice for such tasks because they process inputs sequentially while maintaining a hidden state that summarizes all previously seen elements. However, RNNs suffer from the vanishing/exploding-gradient problem which hampers long-range data dependencies, precisely the situation of which the rotation angle of the $i$-th qubit influences that of the $j$-th qubit for $j>i$. The Long Short-Term Memory (LSTM) model \cite{hochreiter-1997} mitigates this issue by augmenting each recurrent cell with gates (input, forget, output) and an internal memory cell, enabling robust capturing of long-term dependencies in sequential data. Therefore, an LSTM model was chosen to tackle the problem of scaling single qubit rotation angles.

\begin{figure}[t]
\centering
\includegraphics[width=0.46\linewidth]{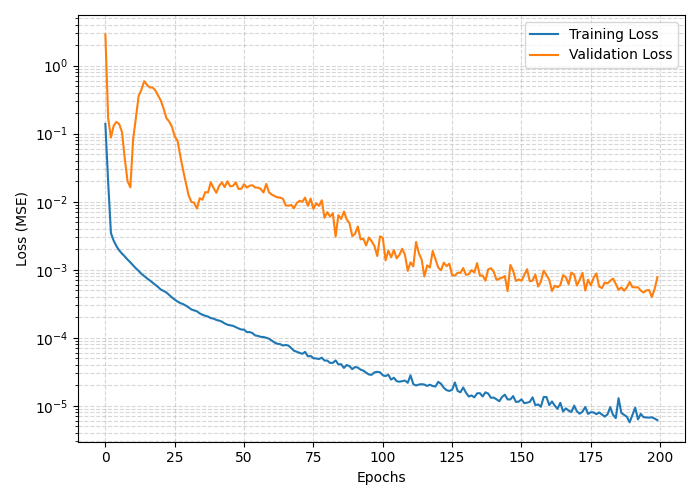}
\includegraphics[width=0.46\linewidth]{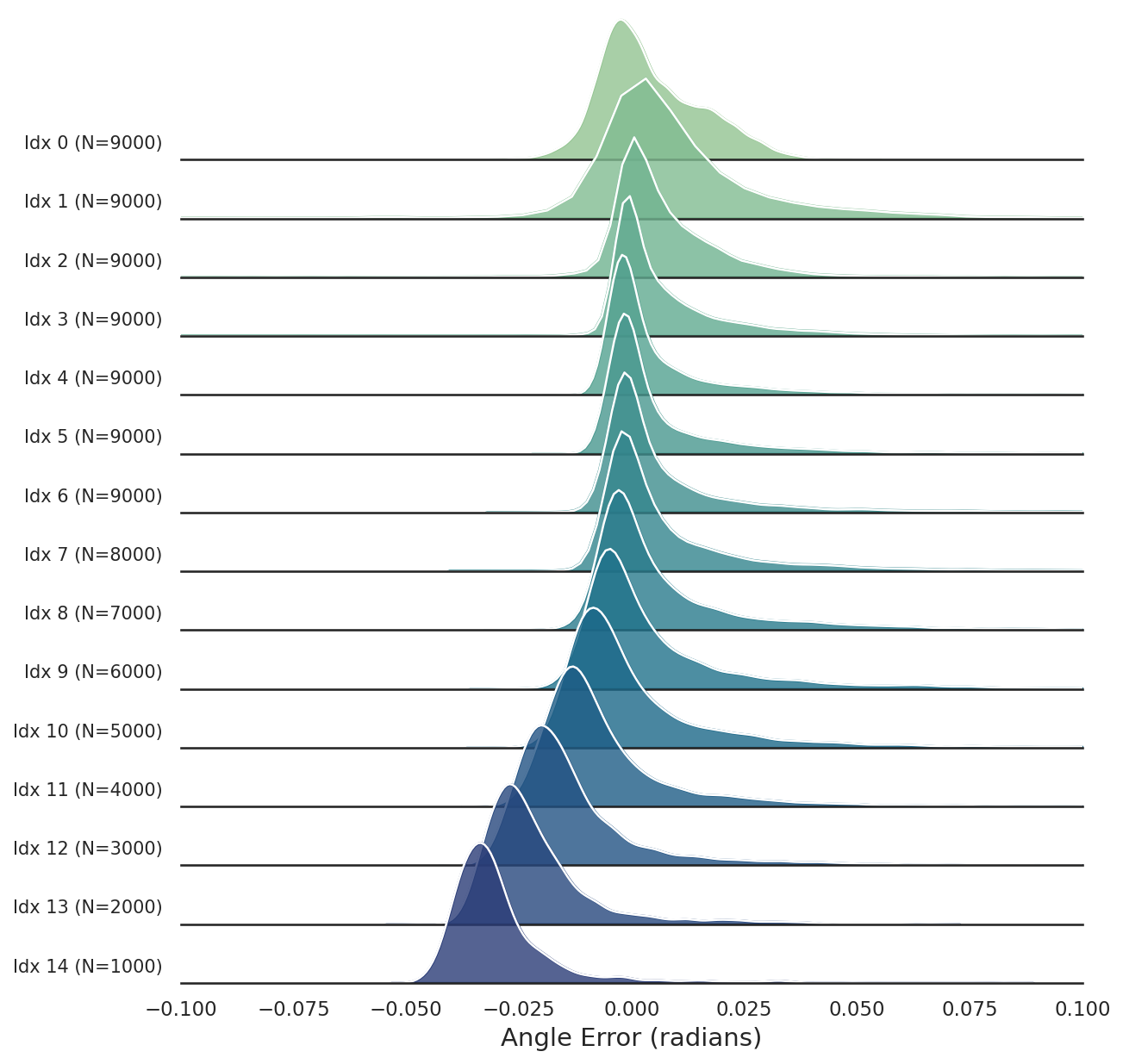}
\caption{\textbf{(Top)} Training and validation loss plot of the LSTM model learning the generalized W state. As mentioned in the text, the model is learning a mapping from $\mathbf{p} \rightarrow \boldsymbol{\theta}$ where $\mathbf{p}$ is a vectorized representation of the generalized W state, and $\boldsymbol{\theta}$ is a vector of single qubit rotation parameters. The model was trained on circuit lengths from 2 to 6, and validated on lengths 6 to 15.
\textbf{(Bottom)} All angle error distributions plotted. Here, the LSTM was evaluated on the $\mathbf{p} \rightarrow \boldsymbol{\theta}$ mapping for circuit lengths from 7 to 15 following training of lengths 2 to 6. The mean of each error distribution slightly deviates from $0$ as the circuit length increases, but remains on the order of $10^{-2}$. }
\label{fig:gen_w_state_training}
\end{figure}

\subsection{Quantum circuit Syntax Generalization}

Large language models (LLMs) have repeatedly shown a remarkable ability to internalize the syntactic and grammatical structure of languages \cite{johnsen2025grammaticalityjudgmentshumanslanguage}. Moreover, equipping LLMs with Chain-of-Thought (CoT) greatly improves reliability on complex tasks \cite{wei2023chainofthoughtpromptingelicitsreasoning}. These two properties make a fine-tuned, pre-trained LLM a natural candidate for generalizing the quantum circuit syntax (i.e., the layout, ordering, and choice of quantum gates in a circuit represented in natural language through openQASM). We stress that the LLM does not interact with the rotation angle parameters; rather, the LLM will only generalize the placement and types of quantum gates and physical qubit measurements \& resets. The output of the fine-tuned, pre-trained LLM is an openQASM representation of an HQC ``skeleton," without the rotation angle parameters, see Fig.~\ref{fig:workflow}. 

The prompt-to-answer pipeline is as follows. First, the LLM will be prompted to \textit{``Generate an openQASM skeleton of a holographic quantum circuit preparing the W state for Y qubits"}. Next, the LLM will use CoT to enumerate the logical steps that define the holographic construction. This includes the number of sites, placement of single qubit rotations, placement of entangling gates, and reset \& measurement operations. Since the model is forced to articulate these steps, it is less likely to produce incorrect openQASM code. Finally, following CoT reasoning, the LLM will output the aforementioned openQASM circuit skeleton.

\section{\label{sec:benchmark}Analytically Solvable Benchmark}

We benchmark our hybrid machine learning framework by scaling up the HQC representation of the generalized W state described in Sec.~\ref{sec:w_decomp}. As noted there, this family of states provides simple but non-trivial examples of non-translationally invariant quantum states, all of which can be described by an MPS with the simplest non-trivial bond dimension of 2.  The benchmarking will proceed as follows. The first two subsections describe the training and validation of the LSTM towards scaling the rotation angle parameters for the generalized W state (by generalizing $\mathbf{p} \rightarrow \boldsymbol{\theta}$ mapping) and the standard W state (by extending $\boldsymbol{\theta}$ to longer sequences). The third subsection describes the fine-tuning procedure of a pre-trained LLM with Chain of Thought in producing a openQASM representation of the quantum circuit. Finally, the fourth subsection compares the shot distribution of a generated OpenQASM script for an arbitrary generalized W state against the true distribution using a noiseless quantum circuit simulator.

\subsection{Scaling rotation angles for the generalized W state}

In this section, we describe the application of an LSTM towards scaling up the rotation angles of the generalized W state. Here, the objective of the LSTM is to learn and generalize the mapping $\mathbf{p} \rightarrow \boldsymbol{\theta}$ where the $j$-th element of $\boldsymbol{\theta}$ is:
\begin{equation}\label{eq:p_theta_mapping}
    [\boldsymbol{\theta}]_j = \sin^{-1}\left[\sqrt{\frac{p_j}{\sum_{i \leq j}p_i}}\right] \, ,
\end{equation}
which connects the parameterization of the state (through the vector $\mathbf{p}$) with the parameterization of its HQC (through the rotation angles $\boldsymbol{\theta}$), see Sec.~\ref{sec:w_decomp}.

Next, we describe the procedure of curating the training and validation sets. Here, we uniformly sampled $1,000$ $\mathbf{p}$ vectors per length $L$ for $L \in [2,15]$, normalized each $\mathbf{p}$ such that $\sum_ip_i=1$, then analytically computed the rotation angle vector $\boldsymbol{\theta}$ from each $\mathbf{p}$. We then split the dataset into training ($2$ to $6$ lengths) and validation ($7$ to $15$ lengths).  Padding up to length $15$ was used to account for the variable length with respect to the $\mathbf{p}$ and $\boldsymbol{\theta}$ vectors. A 2 layer, bidirectional LSTM with hidden size of 64 was employed to learn this mapping. A full description of the model and training hyper parameters can be found in Appendix~\ref{apdx:hyperparameters}. The model's loss function, computed only on the unmasked terms, is the mean squared error (MSE):
\begin{equation}\label{eq:MSE}
    \mathcal{L}= \frac{1}{N}\sum_{i=0}^N(\theta_i - \hat{\theta}_i)^2
\end{equation}
Here, $\theta_i$ is the LSTM-generated $i$-th rotation angle and $\hat{\theta}_i$ is the true $i$-th rotation angle. 

We now describe the results following training. The top panel of Fig.~\ref{fig:gen_w_state_training} shows the evolution of the training and validation loss over the course of training. The model converges to a final training loss on the order of $10^{-5}$ and a final validation loss on the order of $10^{-3}$, indicating that the model does not over-fit despite the increase in sequence length.  The bottom panel of Fig.~\ref{fig:gen_w_state_training} shows the final per-site angle error distribution, calculated as $\theta_i - \hat{\theta}_i$, over the entire validation set of $7$ to $15$ qubits. The error mean slightly deviates from $0$ as the circuit length increases, but remains on the order of $10^{-2}$. Such a result confirms that the LSTM robustly reproduces the analytical mapping for sizes never seen during training. 

The empirical results demonstrate the following: first, the LSTM has implicitly captured the function dependence encoded in Eq.~\ref{eq:p_theta_mapping}, as evidenced in the low final validation loss. Second, despite only being trained on mappings up to $6$ lengths, the LSTM provides accurate predictions up to $15$ lengths without any architectural change. This key result means that the rapid scale-up of HQC parameters for much larger generalized W states can be achieved by training a model on small-sized examples. 

In summary, based on the training and validation loss plots and the angle error distribution plot, we conclude that the LSTM successfully learned and generalized the mapping $\mathbf{p} \rightarrow \boldsymbol{\theta}$. This trained surrogate will provide the rotation angle parameters to be combined with an LLM-generated openQASM skeleton. 

\subsection{Scaling rotation angles for the standard W state}

\begin{figure}[t]
\centering
\includegraphics[width=0.48\linewidth]{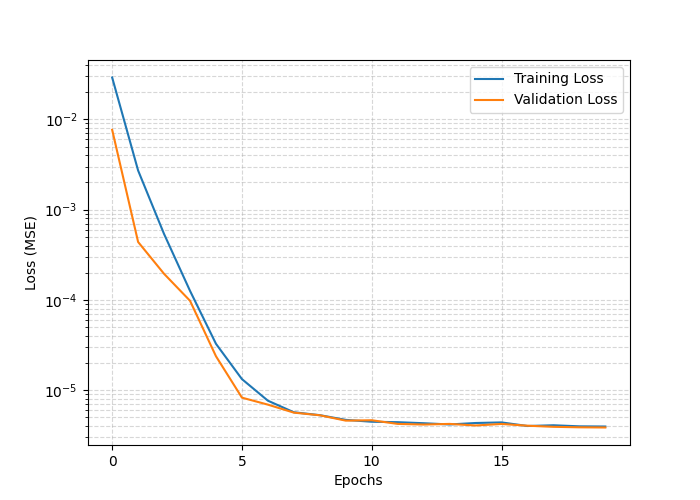}
\includegraphics[width=0.48\linewidth]{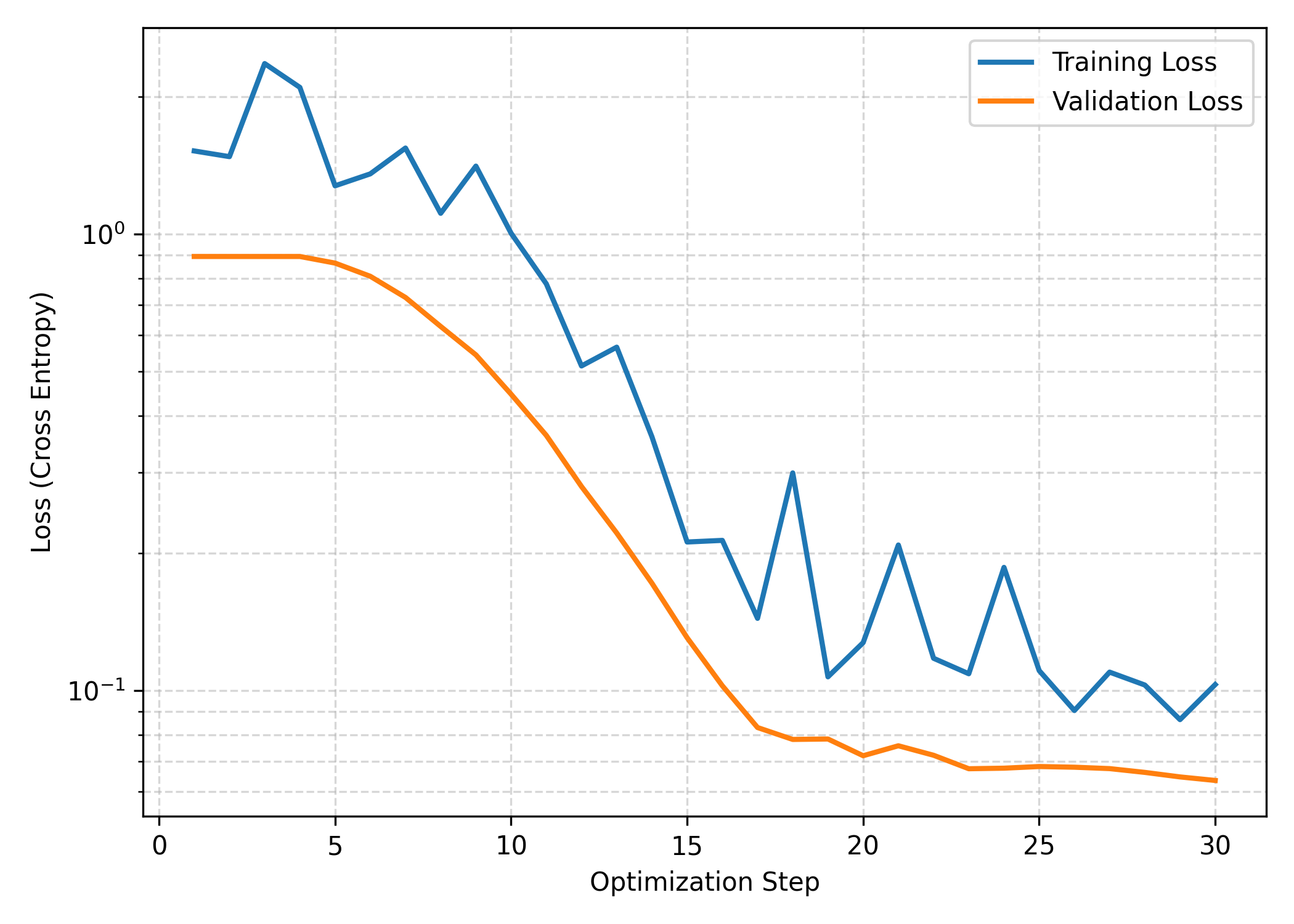}
\caption{\textbf{(Left) }Training and validation loss plot of LSTM model learning the single qubit rotation angles for the standard W state. The model is learning to extend a single qubit rotation angle vector $\boldsymbol{\theta}$. The model was trained on circuit lengths from 2 to 6, and validated on lengths of 6 to 15. \textbf{(Right)} Training and validation loss plot of fine-tuning the LLM towards outputting the skeleton of an openQASM script representation of a holographic quantum circuit. }
\label{fig:loss_standard_w_state_llm}
\end{figure}

In this section, we describe the application of the LSTM towards generalizing the standard (uniform) W state. Here, the objective of the LSTM is different than that of the generalized W state, since there is no reference to the vector $\mathbf{p}$ but only to the rotation angle vector $\boldsymbol{\theta}$. Now, the LSTM is tasked to extend an input rotation angle vector $\boldsymbol{\theta}$ corresponding to an $n$-qubit state to a new rotation angle vector $\boldsymbol{\theta}$ for any state with $m > n$ qubits. Here, the $j$-th element of $\boldsymbol{\theta}$ is:

\begin{equation}\label{eq:w_state_theta}
    [\boldsymbol{\theta}]_j = \sin^{-1}\left[\sqrt{\frac{1}{j+1}}\right] 
\end{equation}

Next, we describe the procedure of curating the training and validation sets. The vector $\boldsymbol{\theta}$ was computed using 1,000 uniformly spaced values of $j$ over the interval $[2,15]$. Such a construction creates a continuous analog of $\boldsymbol{\theta}$ and artificially produces more training examples for the LSTM. The standard W state dataset was split into training (length of 2 to 6) and validation (length of 7 to 15). In order to present the LSTM with a fixed-length input, the training dataset was broken into overlapping windows of length $60$. Identical to the learning of the generalized W state, a 2 layer, bidirectional LSTM with hidden size of 64 was employed to learn this rotation angle extension problem. The model loss function was computed using MSE, shown in Eq.~\ref{eq:MSE}. A full description of the model and training hyper parameters can be found in Appendix~\ref{apdx:hyperparameters}.

Finally, we describe the performance of the trained LSTM. The evolution of the training and validation loss during training can be seen in Fig.~\ref{fig:loss_standard_w_state_llm}(right), with a final training and validation loss both on the order of $10^{-5}$. The final losses indicate that the model has learned the analytical relationship shown in Eq.~\ref{eq:w_state_theta}, and can extrapolate it to unseen circuit lengths. We conclude that training an LSTM on rotation angle vectors up to lengths of $6$ suffices to generate correct angles for circuit lengths as large as $15$. This result implies that a modestly sized training set can be leveraged to obtain HQC parameters for arbitrarily large standard W states.

In summary, we conclude that the LSTM has successfully learned to extend the rotation angle vector $\boldsymbol{\theta}$ of a standard W state from a small system to a much larger-sized system.

\subsection{OpenQASM script generation}

% \begin{figure}[t]
% \centering
% \includegraphics[width=0.4\linewidth]{imgs/finetune_LLM.png}
% \begin{lstlisting}[basicstyle=\ttfamily\scriptsize]
% OPENQASM 2.0;
% include "qelib1.inc";
% qreg q[2];
% creg c[11];
% // BEGIN FIRST SITE
% ry() physical;
% x physical;
% cx physical, ancilla;
% x physical;
% measure physical;
% reset physical;
% // END FIRST SITE
% // BEGIN SITE (REPEATED 10 TIMES)
% cx physical, ancilla;
% ry() physical;
% cx physical, ancilla;
% rz() physical;
% rz() ancilla;
% cx physical, ancilla;
% ry() physical;
% cx physical, ancilla;
% rz() physical;
% rz() ancilla;
% measure physical;
% reset physical;
% // END SITE
% \end{lstlisting}
% \caption{\textbf{(Top):} Training and validation loss plot of fine-tuning the LLM towards outputting the skeleton of an openQASM script representation of a holographic quantum circuit. 
% \textbf{(Bottom):} Example output from the fine-tuned LLM for an openQASM script corresponding to a W state of length 11. The intermediate site preparation and measurement is repeated 10 times. The rotation angles are intentionally left blank to ensure that the LLM deals only with natural language and not with numerical values. }
% \label{fig:loss_LLM}
% \end{figure}
In this section, we describe the application of a pre-trained Large Language Model towards generating skeletons of openQASM representations of holographic quantum circuits. By skeleton, we refer to an openQASM script that contains the type, position, and ordering of all quantum gates (single-qubit gates or two-qubit gates), but leaves all the rotation parameters blank. The rotation parameters will be supplemented by the previously discussed LSTM model. The hyperparameters of the model and the training procedure can be found in Appendix~\ref{apdx:hyperparameters}. 

The base model was "Llama-3.1-70B-Instruct", obtained from Hugging Face's public repository. Prior to fine-tuning, the model was quantized using the BitsAndBytes library (4-bit, NF4 format) to reduce the memory consumption of our Graphics Processing Unit (GPU). Parameter-Efficient Fine-Tuning (PEFT) was then employed, in particular Low-Rank Adaptation (LoRA) \cite{hu2021loralowrankadaptationlarge}, which significantly limits the number of trainable parameters while maintaining performance comparable to regular fine-tuning.

The training set consisted of openQASM scripts for W-state HQCs (see Sec.~\ref{sec:w_decomp}) of length 2 to 10. The validation scripts covered HQCs for lengths 11 to 15, ensuring that the model can extrapolate to larger systems. All scripts were tokenized with the in-house tokenization format provided by Llama-3.1. The loss function chosen was Cross Entropy, which is the negative log-likelihood of the true next-token distribution under the model's next token prediction, calculated as:

\begin{equation}
    \mathcal{L} = -\sum_{t=1}^L \log p_{\zeta}(y_t | y_{<t})
\end{equation}

Here, $y_t$ is the token at position $t$ in the target sequence and $p_{\zeta}(y_t | y_{<t})$ is the soft-max probability produced by the model (with parameters $\zeta$). 

During training, the model was allowed to use Chain-of-Thought (CoT) reasoning before producing the final skeleton. Here, the model was trained on an NVIDIA H100 GPU. The top panel of Fig.~\ref{fig:loss_standard_w_state_llm}(left) shows the evolution of the training and validation loss, with both final values on the order of $\sim 10^{-1}$. An example length-11 skeleton displayed by the model is displayed in the bottom panel of Fig.~\ref{fig:qasm_example}. In this example, the identifiers "physical" and "ancilla" represent the physical qubit and the ancilla qubit, respectively. Moreover, the act of measuring the physical qubit into a specific classical register is deliberately omitted to keep the LLM focused on pure syntactic generation and added in post-processing. 

The final training and validation losses are both on the order of $10^{-1}$, indicating that the LLM has learned and generalized the syntax of such openQASM scripts with low error. Moreover, the LLM respects the syntax of openQASM scripts, such as qubit ordering for CNOT gates and measuring and resetting the physical qubit. Therefore, we conclude that the fine-tuned, LoRa-adapted Llama-3.1-70B-Instruct has successfully learned and generalized the skeleton representation of HQCs. 

\subsection{Evaluation}

\begin{figure*}[t]
\centering
\includegraphics[width=\linewidth]{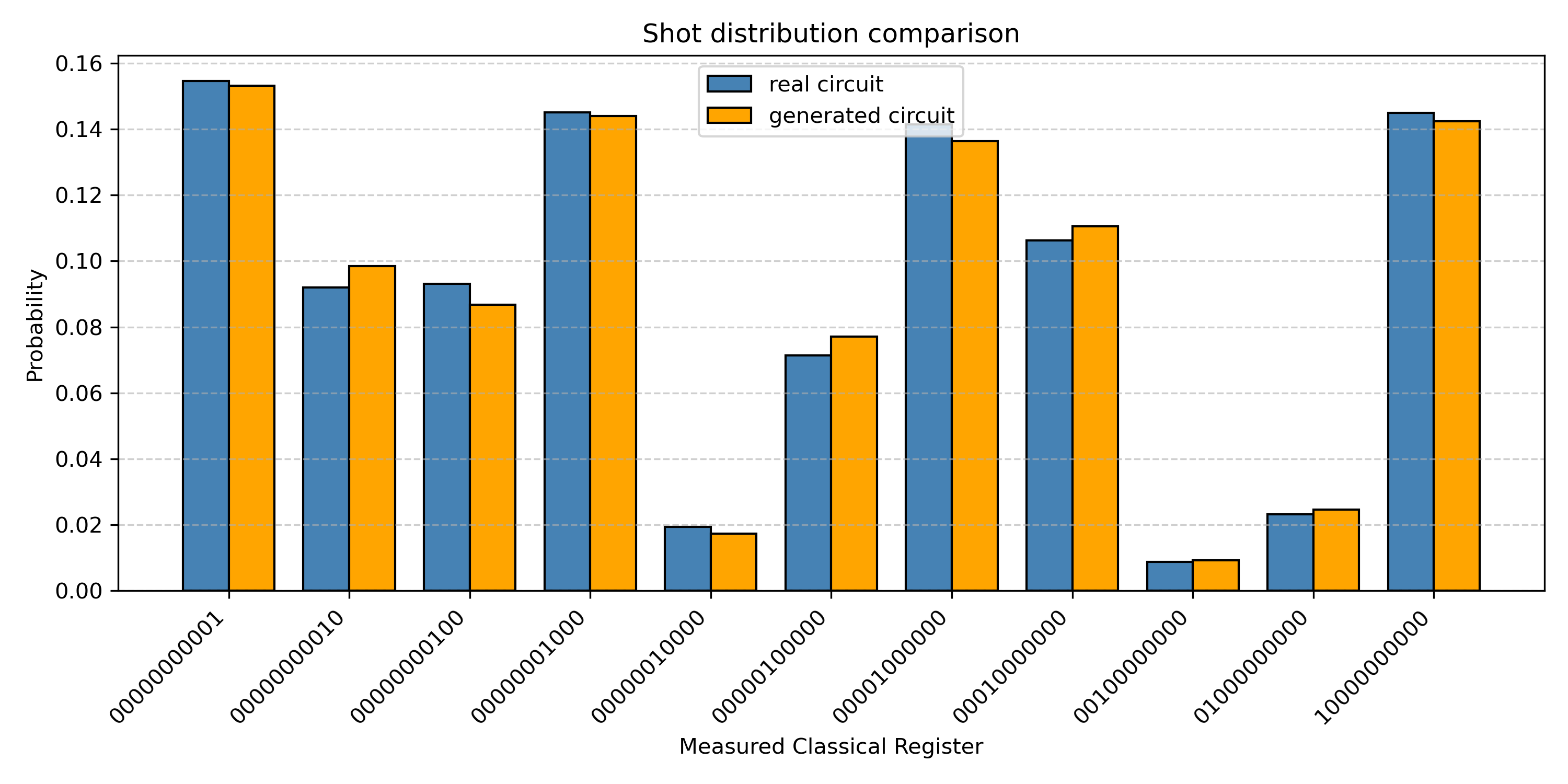}
\caption{Comparison plot of the length-11 relative probability distributions (with $10^5$ shots) of the analytical (blue) generalized W state and the generated (orange) generalized W state produced by combining LSTM generated angles and LLM generated qasm script. Here, the $\mathbf{p}$ vector is (0.146, 0.024, 0.009, 0.104, 0.137, 0.075, 0.018, 0.147, 0.093, 0.0945, 0.153). We observe excellent agreement between these two distributions. The total-variation (TV) distance between the distributions is $1.3\times 10^{-2}$, which can be attributed to finite shot count.  }
\label{fig:shot_distributions}
\end{figure*}

In this section, we detail the end-to-end HQC scale up procedure using our hybrid machine learning model. Here, to generate HQCs for a system of $m>n$ qubits given training data up to $n$ qubits, we combine the outputs of a single qubit rotation angle predictor (LSTM) with the openQASM circuit-skeleton generator (LLM). The complete workflow is illustrated in Fig.~\ref{fig:workflow} and is as follows. 

First, the user will gather either a probability vector $\mathbf{p} \in \mathbb{R}^m$ or a single qubit rotation angle vector for the first $n$ qubits $\boldsymbol{\theta} \in \mathbb{R}^n$. Second, the user will input their vector ($\mathbf{p}$ or $\boldsymbol{\theta}$) into the LSTM, and receive an output of rotation angles $\boldsymbol{\theta} \in \mathbb{R}^m$. Third, the LLM will receive a prompt encoding the system size of $m$ qubits, and will output a skeleton of an openQASM script of $m$ qubits. Fourth, the empty rotation angles in the skeleton are replaced with the values in the newly predicted rotation angle vector $\boldsymbol{\theta} \in \mathbb{R}^m$. After substitution, the script is passed through a post-processor that adds in the measurement of physical qubit values into their respective classical registers. Finally, the resulting openQASM file is loaded into Qiskit~\cite{javadiabhari2024quantumcomputingqiskit} and executed on the Aer Simulator to obtain a shot distribution.

We assessed the fidelity of the end-to-end pipeline with an example generalized W state with length 11. Here, $\mathbf{p}$ is (0.146, 0.024, 0.009, 0.104, 0.137, 0.075, 0.018, 0.147, 0.093, 0.0945, 0.153). The shot distribution of the generated versus analytical openQASM scripts is shown in Fig.~\ref{fig:shot_distributions}, with $10^5$ shots. The total-variation (TV) distance between the two distributions, defined as $\text{TV} = 1/2 \sum_{x \in X} |P(x) - Q(x)|$ where $X$ is the sample space, $P$ is the predicted distribution, and $Q$ is the true distribution, was $1.3\times 10^{-2}$. The non-zero TV value which can be attributed to statistical fluctuations of the chosen shot count. In general, the generated distribution matches the analytical one well, confirming that the hybrid LSTM and LLM architecture has successfully learned both the global structure of the W state and the local rotation parameters required for its preparation. 

\section{Conclusion and Outlook}

In this paper, we proposed a hybrid machine learning framework composed of a Long Short-Term Memory (LSTM) model and a fine-tuned Large Language Model (LLM) used for scaling up holographic quantum circuits (HQCs). Here, each HQC is decomposed into two facets: (1) parameters (e.g., single qubit rotation angles) and (2) syntax (e.g., placement, type, and number of gates). The LSTM model is employed to learn and generalize the parameters, and a fine-tuned pre-trained LLM is trained to learn and generalize the syntax. A high-level overview of our workflow is shown in Fig.~\ref{fig:workflow}. The final openQASM script is obtained by combining the individual outputs of the LSTM and LLM. The LLM chosen for this task was the Llama-3.1-70B-Instruct model. 

We benchmarked this hybrid ML architecture on learning and scaling up HQC representations of a family of generalized W states. Here, the LSTM model successfully scaled up rotation-angle patterns from circuits of size $6$ to circuits of length $15$, while the LoRA-adapted LLM extrapolated the gate-level syntax from length $10$ to $15$. For an example generalized W state for length 11, the generated shot distribution (with $10^5$ shots) differed from the exact distribution by a total-variation (TV) distance of $1.3\times 10^{-2}$. Such an error can be explained by the shot noise from a finite number of samples. Consequently, the hybrid architecture can scale the full HQC (with respect to both the parameters and syntax) to system sizes that exceed the training regime by more than a factor of two.

Our results provide a proof of principle that a hybrid architecture of syntax generation and parameter extrapolation can produce scaled HQCs in the simple but non-trivial setting of the generalized W state.  Looking forwards, future research will address the utilization of such architectures to scale more complex HQCs, such as those representing correlated ground states of quantum spin chains, towards the region where HQCs display a scaling advantage over classical tensor networks.  This will involve more complex syntax to be learned as the number of ancilla qubits are increased, as well as the development of techniques to "gauge fix" the ancilla degrees of freedom for different quantum states so that they can be learned consistently in a unified model.

\section{Acknowledgements}
This work was supported by the Office of Naval Research under Contract No.~N00024-22-D-6404. We would like to thank Matt Abernathy and William Daniels for helpful discussions.

\printbibliography
\newpage 

\appendix

\section{\label{apdx:derivation}Derivation of the holographic compilation of the generalized W state}

In this appendix section, we construct analytic construction of holographic algorithms for its construction on two qubits.  Namely, the generalized W state on $n$ qubits is defined as
\begin{align}
|W\left(\mathbf{p}\right)\rangle&=\sum_{i=1}^{n}\sqrt{p_i}|00\dots 1_i\dots 00\rangle\, ,
\end{align}
in which $\mathbf{p}$ is a vector of positive real numbers whose elements sum to 1.  This state can be interpreted as an amplitude encoding of a probability distribution over one-hot vectors, which is to say that a measurement of $|W\left(\mathbf{p}\right)\rangle$ in the computational basis will result in the qubit $i$ being in state 1 and all other qubits being zero with probability $p_i$.  The standard W state amounts to taking $\mathbf{p}=\mathbf{1}/n$, which is an equal weight superposition of one-hot vectors.  We note that our construction below can also be straightforwardly generalized to include an arbitrary phase factor on each one-hot state, but this is inessential for our purposes and so we don't detail the construction here.

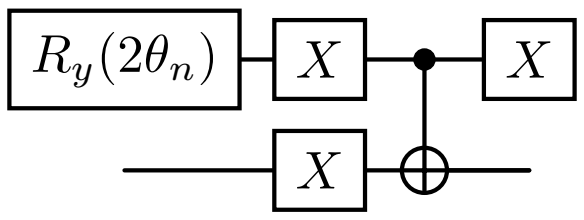
\begin{figure}[h]
\centering
\begin{quantikz}[column sep=0.23cm,row sep=0.15cm]
\gate{R_y(2\theta_{n})} & \gate{X} & \ctrl{1} & \gate{X} \\
\qw & \gate{X} & \targ{} & \qw 
\end{quantikz}
\caption{Circuit decomposition of $U_{n}$ when preparing the generalized W state. The top line is the physical qubit and the bottom line is the ancilla qubit.}
\label{fig:l_n_1}
\end{figure}

The generation of the generalized W state via a holographic algorithm follows from its representation as an MPS with bond dimension 2, the procedure to prepare an MPS sequentially, and an explicit embedding of the sequential preparation isometries into unitary operations~\cite{PhysRevResearch.3.023010}.  To wit, the MPS representation of the generalized W state is
\begin{align}
    |W\left(\mathbf{p}\right)\rangle&=\sum_{i_1\dots i_n} \left(\begin{array}{cc} 1&0\end{array}\right)\mathbb{A}^{[1]i_1} \dots \mathbb{A}^{[n]i_n}\left(\begin{array}{c} 0 \\ 1\end{array}\right)|i_1\dots i_n\rangle\, ,
\end{align}
in which
\begin{align}
    \mathbb{A}^{[j]i_j}&=\left(\begin{array}{cc} \delta_{i_j,0}&\sqrt{p_j}\delta_{i_j,1} \\ 0&\delta_{i_j,0}\end{array}\right)\, .
\end{align}
Using the gauge freedom of MPSs, this can be expressed in left-canonical form~\cite{SCHOLLWOCK201196} as
\begin{align}
    \label{eq:isoprep}|W\left(\mathbf{p}\right)\rangle&=\sum_{i_1\dots i_n} \left(\begin{array}{cc} 1&0\end{array}\right)\mathbb{L}^{[1]i_1} \dots \mathbb{L}^{[n]i_n}\left(\begin{array}{c} 0 \\ 1\end{array}\right)|i_1\dots i_n\rangle\, ,
\end{align}
where
\begin{align}
   \label{eq:Liso} \mathbb{L}^{[j]i_j}&=\left(\begin{array}{cc} \delta_{i_j,0}&\sin\theta_j\delta_{i_j,1} \\ 0&\cos\theta_j\delta_{i_j,0}\end{array}\right)\, ,\\
    \sin\theta_j&=\sqrt{\frac{p_j}{\sum_{i\le j}p_i}}\, ,\cos\theta_j=\sqrt{\frac{\sum_{i<j}p_i}{\sum_{i\le j}p_i}}\, .
\end{align}
Note that each of the tensors $L^{[j]}$ satisfies the left-canonical constraint $\sum_{\alpha i}L_{\alpha \beta}^{[j]i} L_{\alpha \beta'}^{[j]i}=\delta_{\beta\beta'}$, which means that they can be considered as left-isometries when reshaped as $4\times2$ matrices.  This further implies that they can be embedded into unitary operations $\hat{U}_L^{[j]}$ as
\begin{align}
L^{[j]i}_{\alpha \beta} = \langle i\alpha |\hat{U}_L^{[j]}|0\beta\rangle\, ,
\end{align}
in which $|i\alpha\rangle$ denotes a state in the Hilbert space of a physical qubit (index $i$) and an ancilla (i.e., bond) qubit (index $\alpha$).  Using these unitaries, the state can be constructed as
\begin{align}
    \label{eq:Wdecouple}\left(\hat{U}^{[1]}_L |0_1\rangle \dots \left(\hat{U}_L^{[n-1]} |0_{n-1}\rangle \left(\hat{U}_L^{[n]} |0_n 0\rangle\right)\right) \dots\right)&=|W\left(\mathbf{p}\right)\rangle|0\rangle\, .
\end{align}
\begin{figure}[h]
\centering
\begin{quantikz}[column sep=0.23cm,row sep=0.15cm]
\ctrl{1} & \gate{R_y(\theta_j)} & \ctrl{1} & \gate{R_z(\pi/2)} &
\ctrl{1} & \gate{R_z(\theta_j)} & \ctrl{1} & \gate{R_z(-\pi/2)} \\
\targ{} & \qw & \targ{} & \gate{R_z(\pi/2)} &
\targ{} & \qw & \targ{} & \gate{R_z(-\pi/2)}
\end{quantikz}
\caption{Circuit decomposition of $U_{j}$ for $j < n$ when preparing the generalized W state. The top line is the physical qubit and the bottom line is the ancilla qubit. }
\label{fig:l_i}
\end{figure}
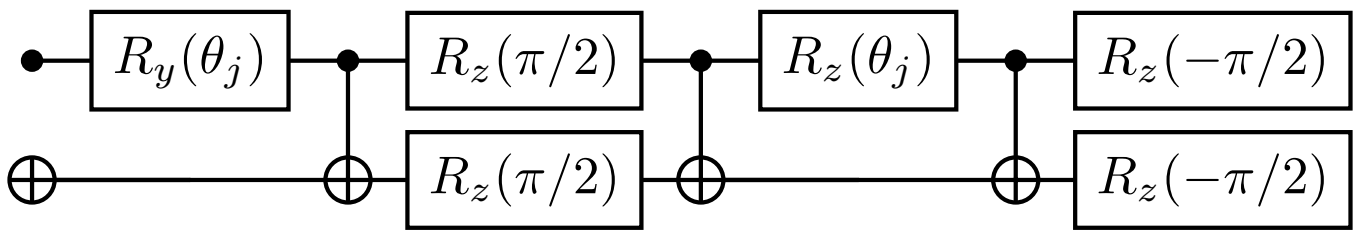
This should be read as a sequential preparation scheme in which each unitary $\hat{U}^{[j]}_L$ is applied to a physical qubit $|i_j\rangle$ in the zero state and the current state of the ancilla register.  At the end of these applications, the physical register holds the generalized W state and the ancilla qubit decouples, as in Eq.~\ref{eq:Wdecouple}.  Because of the sequential nature, we can also generate samples from $|W\left(\mathbf{p}\right)\rangle$ using only a single physical qubit and a single ancilla qubit by measuring and resetting the physical qubit after application of each $\hat{U}^{[j]}$, i.e., as a holographic quantum circuit.  It can be readily verified that the isometries Eq.~\ref{eq:Liso} can be embedded into unitaries given by the gate sequences in Fig.~\ref{fig:l_i} for $U_j$ and Fig.~\ref{fig:l_n_1} for $U_{n}$~\cite{PhysRevResearch.3.023010}.

This representation is not the shortest possible; since the state is real-valued and this sequential circuit is comprised of two-qubit operations, each of the unitary operations $\hat{U}^{[j]}$ is in fact an element of the special orthogonal Group in four dimensions (also called SO(4)) and so can be represented with two Controlled-NOTs (CNOT), six fixed single-qubit rotations, and two parameterized special unitary of dimension two (also called SU(2)) gates~\cite{PhysRevA.69.032315}. However, the gates in Fig.~\ref{fig:l_i} and Fig.~\ref{fig:l_n_1} make explicit the relationship between the parameter $\mathbf{p}$ and the quantum circuit, and so we use it for our analysis.  Finally, as mentioned earlier, the W state is interesting among MPS representations in that the state itself is translationally invariant, but its minimal bond dimension representation is not~\cite{perezgarcia2007matrixproductstaterepresentations}.  This is ideal for our purposes, as it provides a non-trivial but still tractable learning problem for length generalization of holographic quantum circuits.

\clearpage

\section{\label{apdx:hyperparameters}Hyperparameters}

\begin{table}[h]
    \centering
    \caption{Model Hyperparameters for generalized W state}
    \label{tab:hyperparameters_gen_w_state}
    \begin{tabular}{ll}
        \toprule
        \textbf{Hyperparameter} & \textbf{Value} \\
        \midrule
        Learning rate           & $10^{-3}$ \\
        Batch size              & 128 \\
        Number of epochs        & 200 \\
        Optimizer               & Adam \\
        Weight decay            & 0 \\
        Dropout rate            & 0.0 \\
        Hidden layer size        & 64 \\
        Number of layers         & 2 \\
        Activation function       & Tanh \\
        \bottomrule
    \end{tabular}
\end{table}

\begin{table}[h]
    \centering
    \caption{Model Hyperparameters for standard W state}
    \label{tab:hyperparameters_standard_w_state}
    \begin{tabular}{ll}
        \toprule
        \textbf{Hyperparameter} & \textbf{Value} \\
        \midrule
        Learning rate           & $10^{-3}$ \\
        Batch size              & 64 \\
        Number of epochs        & 20 \\
        Optimizer               & Adam \\
        Weight decay            & 0 \\
        Dropout rate            & 0.0 \\
        Hidden layer size        & 64 \\
        Number of layers         & 2 \\
        Activation function       & Tanh \\
        \bottomrule
    \end{tabular}
\end{table}

\begin{table}[h]
    \centering
    \caption{Model Hyperparameters for fine-tuning}
    \label{tab:hyperparameters_fine_tuning}
    \begin{tabular}{ll}
        \toprule
        \textbf{Hyperparameter} & \textbf{Value} \\
        \midrule
        Base model              & Meta-Llama-3.1-70B-Instruct \\
        Quantization            & 4-bit NF4 (double quantization) \\
        Fine-tuning method      & QLoRA (PEFT) \\
        LoRA rank ($r$)         & 8 \\
        LoRA alpha ($\alpha$)   & 32 \\
        LoRA dropout            & 0.05 \\
        LoRA target modules     & q\_proj, k\_proj, v\_proj, up\_proj, down\_proj \\
        Learning rate           & $10^{-4}$ \\
        Optimizer               & Paged AdamW (8-bit) \\
        Per-device batch size   & 1 \\
        Warmup steps            & 4 \\
        Max steps               & 30 \\
        Max sequence length     & 1024 \\
        \bottomrule
    \end{tabular}
\end{table}

\newpage 

\section{\label{apdx:example_qasm_script}Example generated OpenQASM script}

\begin{figure}[h]
\centering
\begin{lstlisting}[]
OPENQASM 2.0;
include "qelib1.inc";
qreg q[2];
creg c[11];
// BEGIN FIRST SITE
ry() physical;
x physical;
cx physical, ancilla;
x physical;
measure physical;
reset physical;
// END FIRST SITE
// BEGIN SITE (REPEATED 10 TIMES)
cx physical, ancilla;
ry() physical;
cx physical, ancilla;
rz() physical;
rz() ancilla;
cx physical, ancilla;
ry() physical;
cx physical, ancilla;
rz() physical;
rz() ancilla;
measure physical;
reset physical;
// END SITE
\end{lstlisting}
\label{fig:qasm_example}
\caption{Example output from the fine-tuned LLM for an openQASM script corresponding to a W state of length 11. The intermediate site preparation and measurement is repeated 10 times. The rotation angles are intentionally left blank to ensure that the LLM deals only with natural language and not with numerical values.}
\end{figure}
\end{document}